\documentstyle[prl,aps,twocolumn]{revtex}
\input{epsf}

\begin{document}


\title{Longitudinal spin-fluctuations and superconductivity in
ferromagnetic ZrZn$_{2}$ from {\em ab~initio} calculations}

\author
{G. Santi$^{1}$, S.~B. Dugdale$^{1,2}$, T. Jarlborg$^{2}$}

\address{1. H.~H. Wills Physics Laboratory, University of Bristol, Tyndall 
Avenue, Bristol BS8 1TL, United Kingdom}

\address{2. D\'epartement de Physique de la Mati\`ere Condens\'ee,
Universit\'e de Gen\`eve, 24 Quai Ernest Ansermet, CH-1211 Gen\`eve 4,
Switzerland} 

\date{\today}
\maketitle


\begin{abstract}

The recent discovery of superconductivity coexisting with weak itinerant
ferromagnetism in the $d$-electron intermetallic compound ZrZn$_{2}$
strongly suggests spin-fluctuation mediated superconductivity.  {\it
Ab initio} electronic structure calculations of the Fermi surface and
generalized susceptibilities are performed to investigate the viability of
longitudinal spin-fluctuation-induced spin-triplet superconductivity in the
ferromagnetic state.  The critical temperature is estimated to be of the
order of 1~K.  Additionally, it is shown that in spite of a strong
electron-phonon coupling ($\lambda_{\rm ph}=0.7$), conventional $s$-wave
superconductivity is inhibited by the presence of strong spin-fluctuations.

\end{abstract}



The generalization of the Bardeen-Cooper-Schrieffer (BCS) theory to
electron-electron interactions by Kohn and Luttinger \cite{kohn:65}, paved
the way for speculation about the possibility of non-$s$-wave, or
``unconventional'' superconductivity.  Following the suggestion that a
magnetically mediated interaction plays an important role in the
superfluidity of liquid $^{3}$He \cite{brueckner:60}, the search began for
superconductivity in nearly magnetic metals where strong spin fluctuations
might provide the pairing mechanism \cite{berk:66,fay:80}.  Recent
experiments on Sr$_{2}$RuO$_{4}$ \cite{luke:98} have made it a strong
candidate for exhibiting spin-triplet, possibly $p$-wave
superconductivity. For a spin-singlet Cooper pair, where the electrons have
anti-parallel spins, the presence of ferromagnetic spin-fluctuations will
be antagonistic towards the development of such a superconducting state.
However, the recent reports of the coexistence of ferromagnetism with
superconductivity in UGe$_{2}$ \cite{saxena:00} and ZrZn$_{2}$
\cite{pfleiderer:01} suggest a spin-triplet Cooper pairing, probably driven
by such spin-fluctuations. Moreover, in ZrZn$_{2}$, the disappearance of
superconductivity at the same point as magnetism, and the sensitivity of
its occurrence to sample purity \cite{pfleiderer:01,foulkes:77} are perhaps
the strongest indications yet that the superconductivity is intimately
connected with the magnetism in this material.

Unlike other ``magnetic'' superconductors (e.g. borocarbides
\cite{canfield:98} , RuSr$_2$GdCu$_{2}$O$_{8}$ \cite{tallon:99}) where the
magnetism and superconductivity occur in different parts of the unit cell,
in both UGe$_{2}$ and ZrZn$_{2}$ it is the same itinerant electrons that
are thought to form the Cooper pairs as well as produce
ferromagnetism. Moreover, whereas some questions regarding the itineracy of
5$f$ electrons and the roles of the strong magnetocrystalline anisotropy
and quasi-two-dimensional electronic structure can be raised with respect
to UGe$_{2}$, ZrZn$_{2}$ is a three-dimensional intermetallic compound free
of such effects.  Discovered by Matthias and Bozorth \cite{matthias:58} in
the 1950s, it was initially of interest because of the presence of weak
ferromagnetism, in spite of the fact that neither constituent was
ferromagnetic. It has the C15 cubic Laves crystal structure, with a lattice
constant of 7.393\AA~(13.97 a.u.) \cite{pfleiderer:01}. The possibility of
there being triplet pairing in high-purity C15 compounds like TiBe$_2$ and
ZrZn$_2$ was first suggested by Fay and Appel \cite{fay:80}. In this Letter
we investigate the viability of their suggestion in the case of ZrZn$_{2}$.

\begin{figure}
\begin{center}
\epsfxsize=210pt
\epsffile{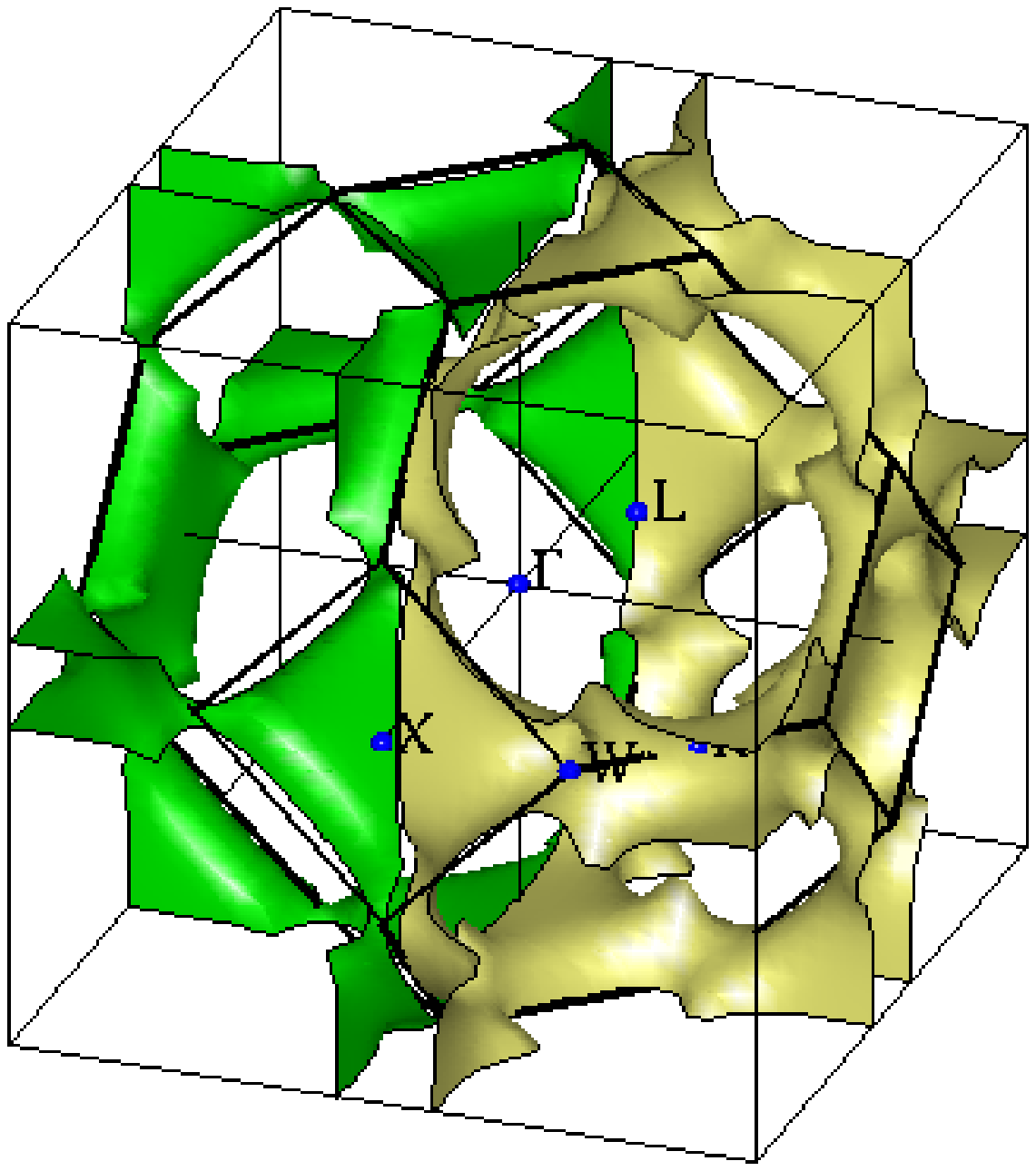}
\epsfxsize=210pt
\epsffile{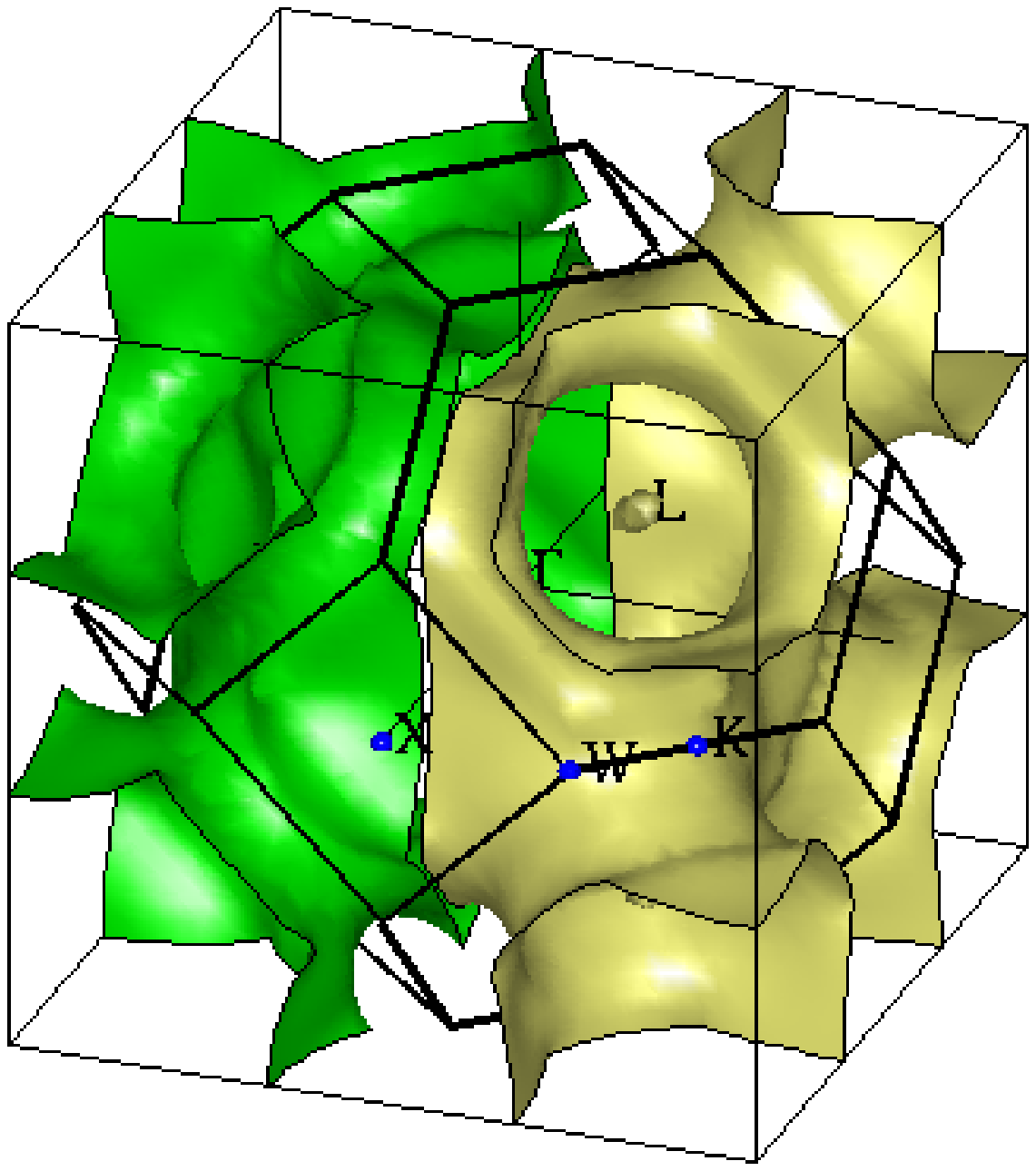}
\epsfxsize=210pt
\epsffile{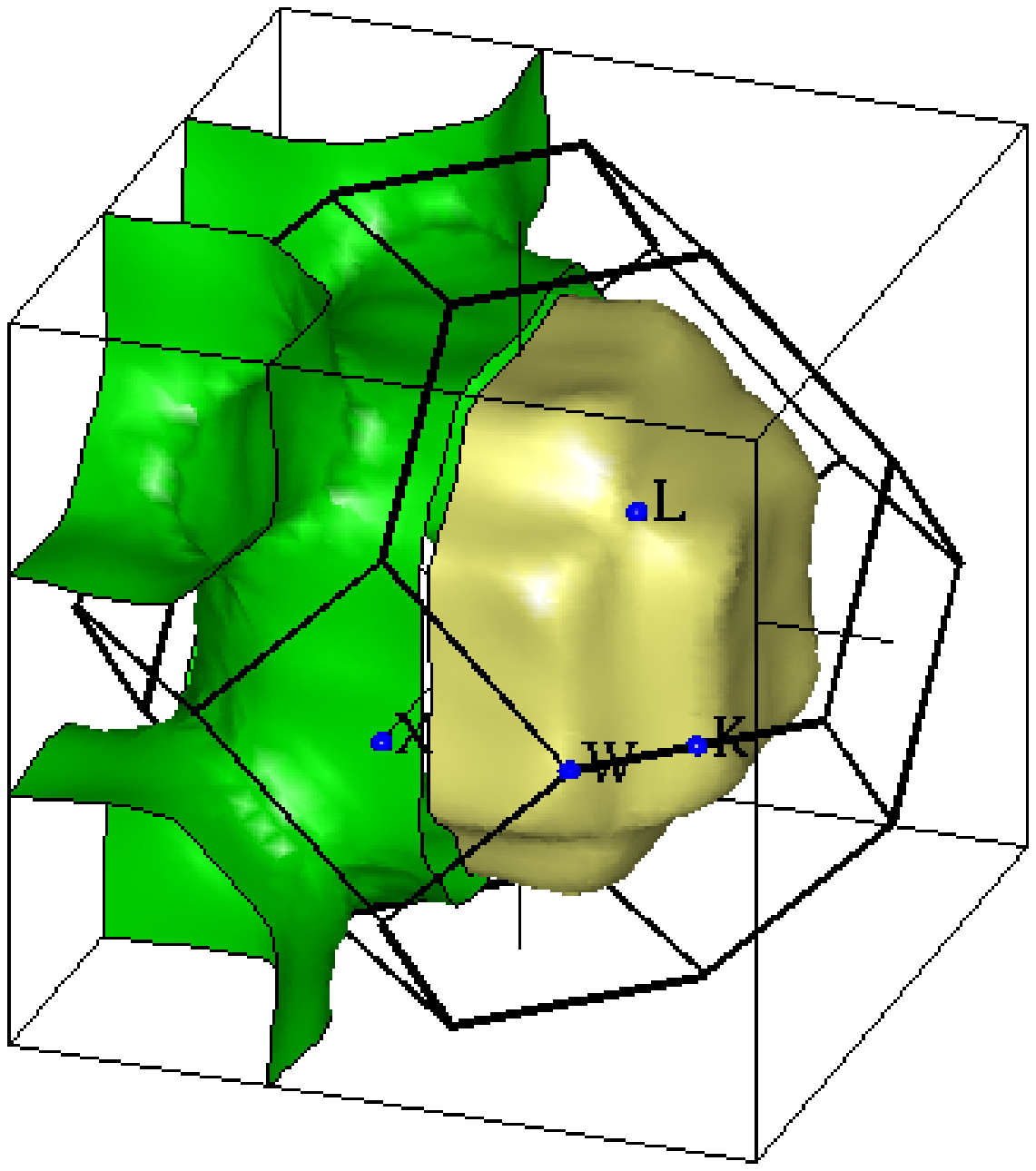}
\epsfxsize=210pt
\epsffile{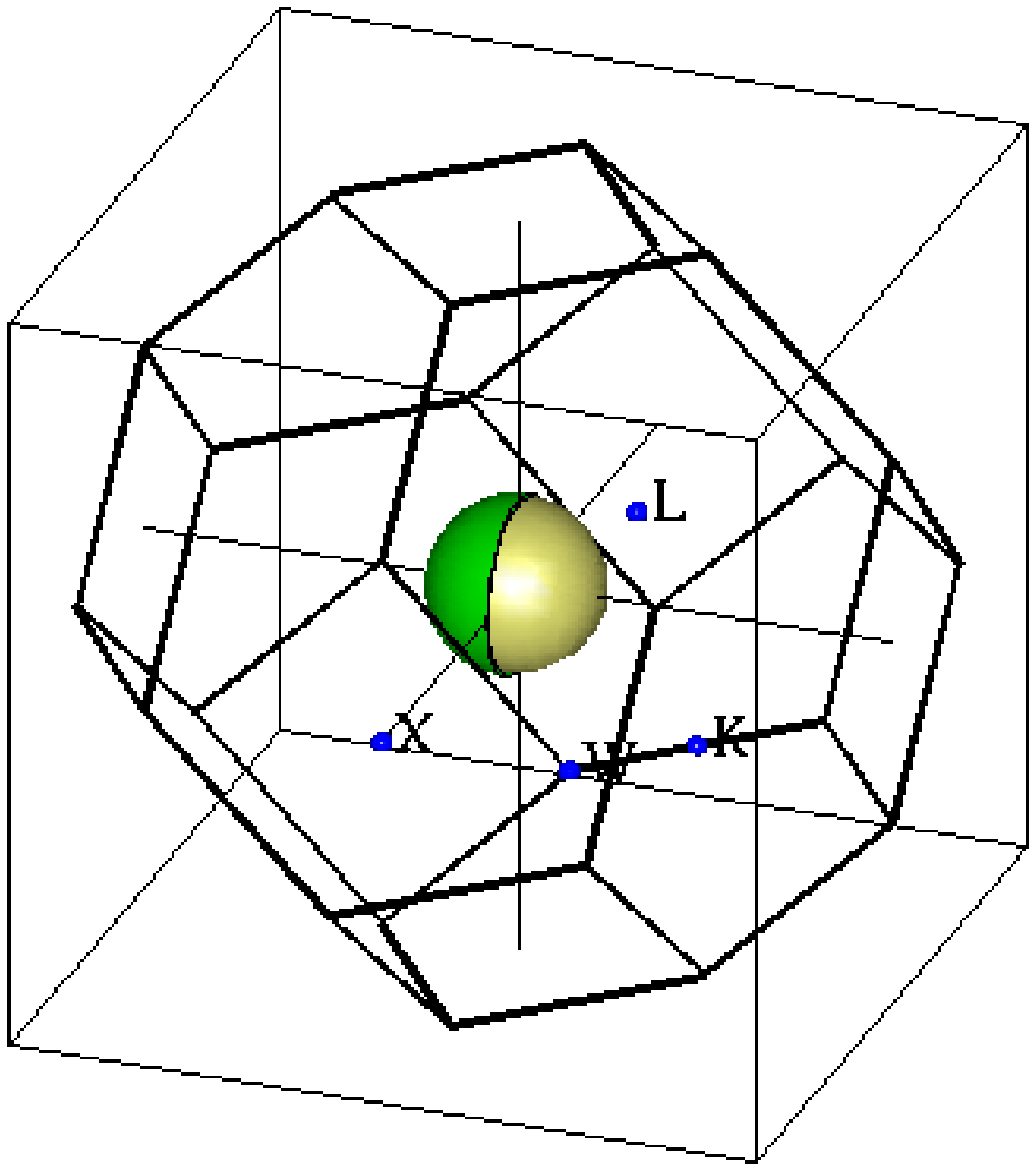}
\end{center}
\caption{
\label{fs}
The spin-polarized Fermi surface for $a = 13.6$ a.u., from bands 27---30
(top to bottom). The majority spin ($\uparrow$) sheets are shown on the left
hand side and the minority spin ($\downarrow$)  on the right. }
\end{figure}

Firstly, we have calculated the electronic structure of ZrZn$_{2}$ using
the LMTO method \cite{andersen:75}. Exchange-correlation effects are
described within the local spin density approximation (LSDA).
Self-consistency was attained using 505 k-points within the irreducible
wedge of the face-centered cubic Brillouin zone (BZ). The basis included
$s$, $p$, $d$ and $f$ states for all atoms. Our results agree with the
calculations of Jarlborg and Freeman \cite{jarlborg:80} and Huang {\it et
al.}  \cite{huang:88} in the non-magnetic state and those of de Groot {\it
et al.} \cite{degroot:80} and Jarlborg {\it et al.} \cite{jarlborg:81} in
the spin-polarized one. The Fermi surface (FS), comprising four sheets, is
shown in Fig.~\ref{fs}. Both non-magnetic (NM) and ferromagnetic (FM)
calculations were performed at a series of different lattice parameters,
and the magnetic moment was found to disappear near 13.47 a.u., where the
calculated pressure is 45 kbar.  This is just below the total energy
minimum, indicating a calculated equilibrium lattice constant of about 13.6
a.u., in excellent agreement with the recent FLAPW calculation of Bruno
{\it et al.} \cite{bruno:01}.  This underestimation ($\sim$2.5\%)
is typical of the LDA, particularly when including $f$ states.  Experiments
confirm that the system is near the ferromagnetic instability, since the
critical pressure for the disappearance of magnetism has been reported in
the range 8.5---22 kbar \cite{smith:71,pfleiderer:01}.

Polarized neutron studies of the magnetization density \cite{pickart:64}
have shown that there is a significant spin-density along the Zr--Zr bond
directions. Our calculation also shows this delocalization, but typically
with 85\% of the total moment on the Zr. The dominance of Zr is also
reflected in the DOS at $E_F$, of which 70\% is of Zr-$d$ character.
Mattocks and Dixon \cite{mattocks:81} inferred an exchange splitting of 4.5
mRy (in a field of 8T) from their de Haas-van Alphen data for orbits on the
$\Gamma$-centered spheroid (band 30). This compares favorably with the
value obtained from the current calculation at the equilibrium lattice
constant (see Table~\ref{elphon}).


Secondly, we have studied the FS and its nesting properties. As seen in
Fig.~\ref{fs}, it is rich in details. Four bands (27--30) cross the Fermi
level. We used the relaxed lattice parameter given by LSDA ($a = 13.6$
a.u.) corresponding to a total moment of 0.17 $\mu_B$/Zr. All these sheets
show a strong Zr-$d$ character, though bands 27 and 30 exhibit a
significant hybridization with Zn-$p$ (25\% and 30\%
respectively). Furthermore, the DOS at $E_F$ is dominated by the
contribution from band 29 (50\%) and band 28 (32\%), whereas band 30
contributes less than 1\%. The effect of the spin-splitting is most
noticeable in the change in the topology of the band 29 sheet, where the
neck near the L point, present for majority spin electrons ($\uparrow$),
disappears for minority ones ($\downarrow$). From Fig.~\ref{fs}, we expect
strong intra and inter-band nesting features, especially along the
$<$100$>$ direction. In order to understand how the nesting will actually
affect the response of the electrons in this system, we have calculated the
bare-band generalized static susceptibility,

\begin{equation}
\label{eq:chi}
\chi^{\sigma\sigma'}_0({\bf q},0) =
\sum_{nn'{\bf k}} \frac{f_{\sigma n{\bf k}}(1-f_{\sigma' n'{\bf k}+{\bf q}})}
{E_{\sigma' n'{\bf k}+{\bf q}} - E_{\sigma n{\bf k}} + i\delta} ,
\end{equation}
where $n$ denotes the band index, $\sigma$ the spin, and $f_{\sigma n{\bf
k}}$ the Fermi-Dirac functions. This sum was calculated on a mesh of more
than $5 \times 10^5$ k-points in the cube (shown in Fig.~\ref{fs}) using a
tetrahedron interpolation technique similar to that of Rath and Freeman
\cite{rath:75}. These calculations were performed for both NM and FM cases
along $<$100$>$, $<$110$>$ and $<$111$>$ in ${\bf q}$-space, at the LSDA
equilibrium lattice constant. All these are peaked at $q = 0$ and (rather
surprisingly, given the FS topology) show very little structure at $q >
0$. This was expected for the NM case because of the ferromagnetism of
ZrZn$_2$, but the absence of finite $q$ peaks in the FM case shows that
this compound does not favor antiferromagnetic (AF) spinwaves as confirmed
by our frozen spinwave calculations. In fact, the only significant $q > 0$
peak is found in the $<$100$>$ direction at $q = 0.2$ (in units of
$2\pi/a$) in $\chi_{0}^{\downarrow\uparrow}(q)$. This peak, originating
from intra-band contributions in band 29 near the corner of the cuboidal FS
sheet (i.e. near the L point) is (obviously) not present
in the NM case. The importance of this peak (and correspondingly the
absence of the other expected ones) is due to the concentration of the DOS
at $E_F$ near the border of the BZ (points X, K and L) where band 29
flattens considerably. In other words, this means that while nesting is
present elsewhere, it is inhibited by the low DOS.


We now turn to the question of longitudinal spin-fluctuation-driven
superconductivity in ZrZn$_{2}$ as proposed by Fay and Appel \cite{fay:80}.
On the FM side of the transition, when the band structure is
different for the two spins, we calculate the longitudinal coupling
constant, $\lambda^{\rm L}_{\rm sf}$, from the generalized
susceptibilities. In the FM region, within the RPA, the pairing potential
can be written as \cite{fay:80}

\begin{equation}
\label{eq:v}
V^{\sigma\sigma}({\bf q}) = 
\frac{I^2({\bf q}) \chi^{-\sigma,-\sigma}_{0{\bf q}}}
{1 - I^2({\bf q}) \chi^{-\sigma,-\sigma}_{0{\bf q}} 
\chi^{\sigma\sigma}_{0{\bf q}}}.
\end{equation}

In the approximation of a spherical FS (not unreasonable, given the dominant
influence of band 29 and its FS topology), the longitudinal coupling
parameter is given by \cite{fay:80},

\begin{equation}
\label{eq:lambda}
\lambda_{{\rm sf},l}^{{\rm L} \sigma} = N_{\sigma}(E_F)
\int_{0}^{2k_{F\sigma}}dq
\frac{q V^{\sigma\sigma}(q)}{2k_{F\sigma}^2}  
P_{l}\left(1-\frac{q^2}{2k_{F\sigma}^2}\right) ,
\end{equation}
where $P_{l}$ is the Legendre polynomial.  The objective is now to estimate
the $s$ and $p$ components of $\lambda_{\rm sf}$ from our band structure
calculation. The exchange integral, $I$, is obtained through the
calculation of the Stoner enhancement, $S$, defined as the {\it increase}
of the exchange splitting of the Zr potential divided by the energy of the
applied magnetic field. The corresponding Stoner factor $\bar{S} = 1 - 1/S$
is simply related to $I$ through $I=\bar{S}/N$, where $N$ is the DOS at
$E_F$. We have calculated $S(q)$ for the FM case (see Table~\ref{elphon})
and for two AF spinwaves ($q=0$ and $q=2\pi/a$).  We find that $S$ is
quickly suppressed ($S(q=2\pi/a) \leq 1.5$) for AF spinwaves. This shows
that ZrZn$_2$ does not support AF fluctuations (which is consistent with
the absence of peaks for $q > 0$ in our calculated generalized
susceptibilities) and that the $q$-dependence of $I$ cannot be
neglected. From $S(q)$, we model $I(q)=I_{0}/(1+b^2q^2)$
\cite{fay:80,mazin:97} with $I_0 = 0.04$ Ry and $b^2 = 0.33$
$(a/2\pi)^2$. Since the contribution to $\lambda_{\rm sf}$ from each
spinwave mode is $\frac{1}{2} S(q) \bar{S}(q)^2$ \cite{jarlborg:86}, we
get, as our first estimate, $\lambda_{\rm sf} = 1.2$ by averaging over
these three modes. We make the further approximation
$\chi_0^{\sigma\sigma}(q) = N_{\sigma}(E_{F})$ $\forall q$, and calculate
the longitudinal $\lambda_{{\rm sf},l}^{{\rm L}\sigma}$ in the $s$ ($l=0$)
and $p$ ($l=1$) channels from Eqs. (\ref{eq:v}) and (\ref{eq:lambda}),
taking a $k_{F\sigma} \approx 0.6$ $2\pi/a$, appropriate to the band 29
sheet. For the relaxed FM case in which $\bar{S} = 1.12$, $\lambda_{\rm
sf}^{\rm L}$ is negligible in the $p$ channel. However, moving closer to
the FM transition, i.e. for $\bar{S} = 1.01$, we get $\lambda_{{\rm
sf},0}^{{\rm L}\sigma} = 1.9, 2.0$ (for $\sigma = \uparrow, \downarrow$) in
the $s$ channel and $\lambda_{{\rm sf},1}^{{\rm L}\sigma} = 0.81, 0.76$
(for $\sigma = \uparrow, \downarrow$) in the $p$ channel.  This shift is
justifiable given that $S$ is extremely sensitive close to the transition.
As noted by Fay and Appel, the $s$ component is much larger than the $p$
one and both diverge when $\bar{S} \rightarrow 1$. These values are
consistent with our previous estimate.

The electron-phonon interaction cannot be ignored and can even be expected
to be rather large owing to the occurrence of conventional
superconductivity in both Zr and Zn, and the large DOS at $E_F$. This
suggests that the electron-phonon coupling could be sufficient to overcome
the pair-breaking effects due to spin-fluctuations. The electron-phonon
coupling constant, $\lambda_{\rm ph}$, can be expressed as $\lambda_{\rm
ph} = \sum_{i}\frac{\eta_{i}}{M_{i} \langle \omega_{i}^2 \rangle}$, where
the sum runs over all atoms, $i$, with masses, $M_{i}$, and phonon
frequencies, $\omega_{i}$, while the numerator, $\eta_{i} = N_{i}(E_{F})
\langle (\nabla V_{i})^2 \rangle$, is the Hopfield parameter that describes
the electronic contribution \cite{dacorogna:84}. Here, $\eta$ was
calculated in the rigid muffin-tin approximation \cite{dacorogna:84},
i.e. retaining only dipolar terms and neglecting electronic screening of
the ionic displacements.  The values for $\langle \omega_{i}^2 \rangle$ are
taken as one half of the Debye frequency of the atom $i$.  Furthermore, we
assume that the volume dependence of $\langle \omega_{i}^2 \rangle$ follows
$\sqrt{aB}$, where $B$ is the (calculated) bulk modulus and $a$ the lattice
parameter, which is reasonable as long as all small-$q$ phonons behave
identically with pressure \cite{pictet:87}.  As shown in Table~\ref{elphon},
$\lambda_{\rm ph}$ is of the order 0.7 near the calculated equilibrium
volume and twice as large at the experimental lattice constant. Ignoring
completely the destructive effects of spin fluctuations, and using $\mu^* =
0.13$ in the McMillan formula \cite{mcmillan:67}, these correspond to
respectable T$_c$'s of about 8 and 21~K, at the respective lattice
constants. The decrease of $\lambda_{\rm ph}$ in the FM state can be
attributed to the smaller total DOS at $E_{F}$ and its pressure dependence
can be ascribed almost entirely to the behavior of the Debye frequency.  In
the vicinity of the ferromagnetic transition, this $\lambda_{\rm ph}$ is
nevertheless insufficient to overcome the dominance of the
spin-fluctuations as indicated by the large $S$ enhancements near the
critical pressure (see Table~\ref{elphon}).  A possibility exists for
phonon-mediated superconductivity at larger pressures, i.e. well outside of
the magnetic region when the Stoner factor would be further decreased to
completely suppress spin fluctuations.  However, our calculations at
$a=13.17$ a.u. (equivalent to 160 kbar) indicate that $\lambda_{\rm ph}$
drops to 0.4 while $S$ is still 4.1 (giving a $\lambda_{\rm sf}$ of the
order of 0.4), which combine to make the conditions non-favorable for
phonon-mediated superconductivity. Note that the persistence of such a
large Stoner enhancement over this range of pressures shows again the
importance of the spin fluctuations and that the large values for $S$ in
the FM region can be related to the observed absence of saturation of the
magnetic moment \cite{pfleiderer:01} and the weak ferromagnetism. Since the
magnetic moment, Stoner enhancement and spin fluctuations are associated
with the Zr sublattice, phonon-mediated superconductivity might be
envisaged to take place within the Zn sublattice, but such an explanation
can be ruled out because the Zn contribution to $\eta_{i}$ is negligible.

Having presented evidence against the possibility of electron-phonon driven
superconductivity, we now try to estimate $T_c$ from the longitudinal spin
fluctuations.  The typical spin-fluctuation cut-off frequency, $\omega_{\rm
sf}$, can be estimated from the Stoner factor by $\omega_{\rm sf}=1/(4NS)$
\cite{mazin:97}, giving about 90~K at the relaxed lattice parameter. Using
the Allen-Dynes formula, we arrive at a simplified expression for the
superconducting transition temperature~:

\begin{equation}
\label{eq:tc}
k_{B}T_{c} = \frac{\hbar \omega_{\rm sf}}{1.2} 
\exp \left(-\frac{1+ \lambda_{\rm ph} +
\lambda^{\rm L,T}_{{\rm sf},0}}{\lambda^{\rm L}_{{\rm sf},1}} \right).
\end{equation}

Note that the rather strong electron-phonon interaction $\lambda_{\rm ph}$
contributes to the mass renormalization (numerator) and is detrimental to
superconductivity in this case. Furthermore, the $s$-wave $\lambda^{\rm
L,T}_{{\rm sf},0}$ contains both the longitudinal (L) and transverse (T)
contributions. From the large measured electronic specific heat coefficient
$\gamma_{\rm exp} = 47$ $\rm mJ\;mol^{-1}K^{-2}$ \cite{pfleiderer:01}, and
our calculated values for $\lambda_{{\rm sf},0}^{{\rm L}\sigma}$ and
$\lambda_{\rm ph}$, we infer a transverse contribution $\lambda_{{\rm
sf},0}^{\rm T}$ of about 0.8. Using the values for the case $\bar{S} =
1.01$, we get a $T_c^{\sigma} = 1.0, 0.8$~K for $\sigma =
\uparrow,\downarrow$ respectively.  These estimates are very approximate
but they confirm, in our opinion, the viability of triplet $p$-wave
superconductivity in ZrZn$_2$.

In conclusion, we have shown that calculations based on our electronic
structure results strongly support the idea that the recently observed
superconductivity in ZrZn$_2$ \cite{pfleiderer:01} is indeed a result of
triplet pairing, as suggested by Fay and Appel \cite{fay:80}.  This would
lead to $p$-wave superconductivity, and since impurity scattering acts as a
pair-breaker for pairing in the $l \neq 0$ channels, the high purity of
samples is crucial.  However, the experimental absence of superconductivity
in the paramagnetic phase just above the critical pressure
\cite{pfleiderer:01} is still unanswered by this theory which predicts an
even larger $T_c$ in the NM region. The answer may lie in the peak of the
transverse susceptibility, $\chi_{0}^{\downarrow\uparrow}(q_{100})$, at
$q=0.2$ which could provide an attractive coupling that would naturally
disappear outside of the FM phase.  Finally, it might be worthwhile
revisiting the properties of C15 compound TiBe$_2$ under pressure since its
electronic structure is very similar to that of ZrZn$_2$ and conventional
superconductivity would be more favored owing to the lighter masses of its
constituents.


We are grateful to S.~M. Hayden for communicating his results prior to
publication and to B.~L. Gy\"orffy and J.~B. Staunton for helpful
discussions.
We also want to acknowledge the financial support of the Swiss National
Science Foundation (GS) and the Royal Society (SBD).



\begin{table}
\caption{Calculated parameters (per formula unit) for various lattice
constants, $a$. Shown are the magnetic moment $\mu$, the exchange
splitting, $\xi$, the Stoner factor, $S$, the density of states at the
Fermi level for the non-magnetic (NM) calculations as well as for the
ferromagnetic (FM) ones in parentheses ($\uparrow$/$\downarrow$), Debye
temperature (set to 370~K at $a = 13.573$ a.u.) used to calculate the
electron-phonon coupling, $\lambda_{\rm ph}$, in both the NM and FM cases,
and the specific heat coefficient renormalized by the electron-phonon
interaction, $\gamma$.}
\label{elphon}
\begin{tabular}{ccccccccc}
$a$ & $\mu$ & $\xi$ & S & DOS(E$_{F}$) & $\theta_{D}$ & $\lambda_{\rm ph}$ &
$\lambda_{\rm ph}$ & $\gamma$ \\ 
a.u. & $\mu_B$ & mRy &  & ${\rm Ry^{-1}}$ & K & (NM) &
(FM) & $\rm mJ/mol K^{-2}$ \\ 
\tableline
13.970 & 0.48 & 19.5 & 2.9  & 68 (18/27) & 265 & 1.42  & 0.90 & 28.6 \\ 
13.573 & 0.10 &  5.0 & 8.3  & 54 (31/25) & 370 & 0.71  & 0.72 & 16.1 \\
13.437 & 0.00 &  0.0 & 9.0  & 52 (26/26) & 420 & 0.56  &  -   & 13.8 \\
\end{tabular}
\end{table}


\end{document}